\documentstyle[twocolumn,epsfig]{jpsj}

\title{Ground State of $\nu=1$ Bilayer Quantum Hall Systems}
\author{Naokazu {\sc Shibata} and Daijiro {\sc Yoshioka}$^1$}
\inst{Department of Physics, Tohoku University, 
Aoba, Aoba-ku, Sendai 980-8578 \\
$^1$Department of Basic Science, University of Tokyo,
Komaba 3-8-1 Meguro-ku, Tokyo 153-8902
}
\recdate{January 27, 2006}

\abst{
The ground-state wave function and the energy gap are calculated
for various layer separations, $d$, and for up to 24 electrons
by the density matrix renormalization group (DMRG) method. 
Two-particle distribution function
and excitonic correlation function are calculated from the ground-state
wave function, 
and the evolution of the ground state with increasing $d$ is analyzed.
The results indicate that the transition is continuous. 
A smooth crossover of the ground
state is found at $d/l \simeq 1.6$ from the excitonic character at small
$d/l$ to the independent Fermi-liquid character at large $d/l$, 
where $l$ is the magnetic length.
}

\kword{bilayer, excitonic state, quantum Hall system, ground state,
phase diagram, two dimensions, density matrix, renormalization group}

\def\runtitle{
Ground State of $\nu=1$ Bilayer Quantum Hall Systems
}
\def\runauthor{Naokazu {\sc Shibata} and Daijiro {\sc Yoshioka}}

\begin{document}
\sloppy
\maketitle
Two-dimensional electron systems under high magnetic field
have many interesting features.
This is typically shown in the 
ground-state phase diagram\cite{Shib1} of monolayer quantum Hall systems:
incompressible liquids\cite{Lagh}, compressible liquids\cite{Jain,Halp}, 
CDW states\cite{Kou1} called stripes, bubbles and Wigner crystals are 
realized only by applying uniform magnetic field. 

In bilayer quantum Hall systems, much more interesting properties 
are expected owing to the additional degrees of freedom of layers.\cite{EM}
The excitonic phase, namely Haplerin's $\Psi_{1,1,1}$ state, 
is one of the ground states realized in
bilayer quantum Hall systems at the total filling $\nu=1$ at a small 
layer separation, $d$, where electrons and holes in different layers
are bound to each other due to strong
interlayer Coulomb interaction.\cite{YMG,YM}
This excitonic state has recently attracted attention
because a dramatic increase in zero bias tunneling conductance 
between the two layers\cite{ZTNC}, and the vanishing of the 
Hall counterflow resistance are observed\cite{CFH1,CFH2}.

As the layer separation is increased, the excitonic phase vanishes, 
and at a sufficiently large separation, the composite-fermion Fermi-liquid 
state is realized in each layer. 
Several scenarios have been proposed for the transition
of the ground state as the layer separation 
increases\cite{HFT,PCF,Mac,Kim,SH,NY,SRM}, 
but how the excitonic state develops into the 
independent Fermi-liquid state has not been fully understood. 
In this paper we investigate the ground state of $\nu=1$ bilayer 
quantum Hall systems using the density matrix renormalization 
group (DMRG) method\cite{DMRG}. 
We calculate the energy gap, two-particle distribution function $g(r)$ 
and excitonic correlation function for 
various values of the layer separation $d$, and show the 
evolution of the ground state with increasing $d$.
The results indicate that the transition of the ground state
from $d=0$ to $\infty$ is continuous. 
There is a smooth crossover of the ground
state at around $d/l \simeq 1.6$ from the excitonic character
to the independent Fermi-liquid character.

The Hamiltonian of the bilayer quantum Hall systems is written as
\begin{eqnarray}
H &=& \sum_{i<j} \sum_{\mib q} V(q)\ {\rm e}^{-q^2l^2/2} 
{\rm e}^{{\rm i}{\mib q} \cdot ({\mib R}_{1,i}-{\mib R}_{1,j})} \nonumber \\
&&+ \sum_{i<j} \sum_{\mib q} V(q)\ {\rm e}^{-q^2l^2/2} 
{\rm e}^{{\rm i}{\mib q} \cdot ({\mib R}_{2,i}-{\mib R}_{2,j})} \nonumber \\
&&+ \sum_{i,j} \sum_{\mib q} V(q)\ {\rm e}^{-qd}e^{-q^2l^2/2} 
{\rm e}^{{\rm i}{\mib q} \cdot ({\mib R}_{1,i}-{\mib R}_{2,j})},
\label{Coulomb}
\end{eqnarray}
where ${\mib R}_{1,i}$ are the two-dimensional guiding center 
coordinates of the $i$th electron in layer-1 and 
${\mib R}_{2,i}$ are those in layer-2. 
The guiding center coordinates satisfy the commutation relation,
$[{R}_{j}^x,{R}_{k}^y]={\rm i}l^2\delta_{jk}$. 
$V(q) =2\pi e^2/q$ is the Fourier transform of the
Coulomb interaction.
We consider uniform positive background charge to cancel the
component at $q=0$.
The wave function is projected onto the lowest Landau level.
We assume zero interlayer tunneling and fully spin-polarized 
ground state.

We calculate the ground-state wave function  
by the DMRG method\cite{DMRG}, which is a real space 
renormalization group method combined with the exact 
diagonalization method.
The DMRG method provides the low-energy eigenvalues and 
corresponding eigenvectors of the Hamiltonian within a 
restricted number of basis states\cite{Shib3,Shib,DY}.
The accuracy of the results is systematically controlled 
by the truncation error, 
which is smaller than $10^{-4}$ in the present calculation.
We investigate systems of various size with up to 
24 electrons in the unit cell keeping 500 basis states in each block.

In Fig.~1 we first show the lowest excitation gap
at the total filling factor $\nu=1$.
The total filling factor is defined by $\nu=N_e/N$,
where $N_e=N_1+N_2$ is the total number of electrons 
and $N$ is 
the number of one-particle states in each layer.
$N_1$ and $N_2$ are the numbers of 
electrons in layer-1 and layer-2, respectively,
and we impose the symmetric condition of $N_1=N_2$.
The gap shown in Fig.~1 is calculated for $N_e=24$ 
and $L_x/L_y=1.6$ in the unit cell $L_x\times L_y$ 
with the periodic boundary conditions in both $x$ and $y$ directions.
The aspect ratio $L_x/L_y$ is chosen from the minimum of 
the ground-state energy with respect to $L_x/L_y$ at around
$d/l=1.8$, where the minimum structure appears in the ground-state
energy.

We can clearly see a large excitation gap for $d/l<1.2$,
where the excitonic ground state
is expected both theoretically and experimentally
\cite{YMG,YM,ZTNC,CFH1,CFH2,HFT,PCF,Mac,Kim,SH,NY,SRM}.
The excitation gap rapidly decreases with increasing  $d/l$
from $1.2$, and it becomes very small for $d/l>1.6$. 
This behavior is consistent with experiment.\cite{Wies}
Although the excitation gap for $d/l> 1.7$ is not presented 
in the figure for $N_e=24$ because of the difficulty in calculating
energy of excited states in large systems, we do not find any sign 
of level crossing in 
the ground state up to $d/l\sim 4$, where two layers are 
almost independent.
These results suggest that the excitonic state at small $d$
continuously crosses over to a compressible state at large $d$.
In the present calculation, it is difficult to conclude whether 
the gap closes at a finite $d/l\simeq 1.6$ in the thermodynamic limit 
or the excitonic state survives with an exponentially small finite gap
even for a large $d/l>1.6$.
We have calculated the excitation gap
in systems of different size and aspect ratio, 
and obtained similar results as shown in the inset of Fig.~1.

In the first excited state, however,  
a level crossing is expected at $d/l\simeq 1.2$,
where we can see a sudden decrease in excitation gap, 
which is due to a crossing of excited states. 
This means the character of the low-energy excitations 
changes at $d/l \simeq 1.2$, which is confirmed 
by a clear change in the correlation functions in the excited 
state as shown later.

\begin{figure}[t]
\epsfxsize=78mm \epsffile{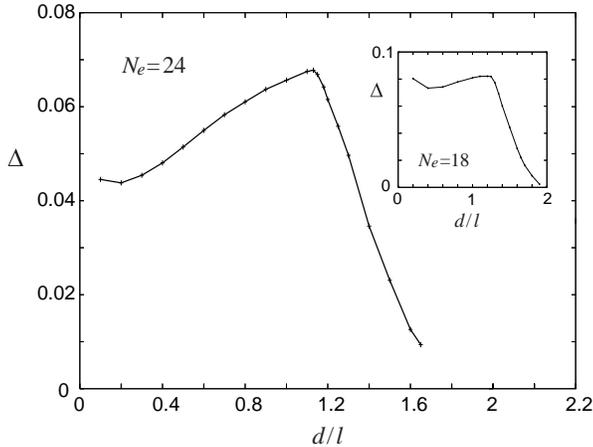}
\caption{
Lowest excitation gap $\Delta$
of bilayer quantum Hall systems at total filling 
factor $\nu=1$. $N_e=24$ and $L_x/L_y=1.6$.
The inset shows the result for $N_e=18$ and $L_x/L_y=1.0$.
The excitation gap $\Delta$ is in units of $e^2/l$.
}
\end{figure}

To study how the excitonic ground state at $d=0$
crosses over to the compressible state at a large $d$, 
we next calculate the exciton correlation defined by 
\begin{equation}
g_{\rm ex}(n) \equiv
\frac{N(N-1)}{N_1N_2}
\langle \Psi | c^\dagger_{1,n} c_{2,n}  
c^\dagger_{2,0} c_{1,0}  |\Psi \rangle,
\end{equation}
where $|\Psi\rangle$ is the ground state and $c^\dagger_{1,n}$ 
($c^\dagger_{2,n}$) is the creation operator of
the electrons in the $n$th one-particle state defined by
\begin{equation}
\phi_n({\mib r})=\frac{1}{\sqrt{L_y\pi^{1/2}l}}
\exp\left\{{\rm i} k_y y - \frac{(x-X_n)^2}{2l^2}\right\}
\end{equation}
in layer-1 (layer-2). 
$X_n=nL_x/N=-k_yl^2$ is the $x$-component of the guiding 
center coordinates and 
$L_x$ is the length of the unit cell in the $x$-direction. 

Because $g_{\rm ex}(n)$ represents the correlations between the 
two excitons created at 
$X=0$ and $X=X_n$, $\lim_{n\rightarrow \infty} g_{\rm ex}(n)\ne 0$ indicates 
the existence of the macroscopic coherence of excitons. 
As shown in Fig.~2, $g_{\rm ex}(n)$ tends to 1 as $d \to 0$.
Indeed Haplerin's $\Psi_{1,1,1}$ state has this macroscopic coherence 
and $g_{\rm ex}(n)=1$ independent of $n$. 
In this figure we have shown $g_{\rm ex}(N/2)$ instead of 
$\lim_{n\rightarrow \infty} g_{\rm ex}(n)$, 
because the largest distance between the two excitons is 
$L_x/2$ in a finite unit cell of $L_x\times L_y$ under the
periodic boundary conditions. 
To check the size effect, we also plot $g_{\rm ex}(N/2-1)$
with a dashed line. Because the difference between 
$g_{\rm ex}(N/2)$ and $g_{\rm ex}(N/2-1)$ is small, we expect 
$g_{\rm ex}(N/2)$ well represents the macroscopic coherence
in the limit of $N\rightarrow \infty$.

With increasing $d$, the excitonic correlation decreases monotonically 
and finally falls down to a negligible value at $d/l \simeq 1.6$. 
Since $g_{\rm ex}(N/2)$ continuously approaches zero, the transition is 
not of the first order. This second order or crossover transition at
around $d/l\simeq 1.6$ is 
consistent with the result of the excitation gap shown in Fig.~1.
The slight difference in the asymptotic form around $d/l\simeq 1.6$ 
is expected to be a finite size effect.

\begin{figure}[t]
\epsfxsize=80mm \epsffile{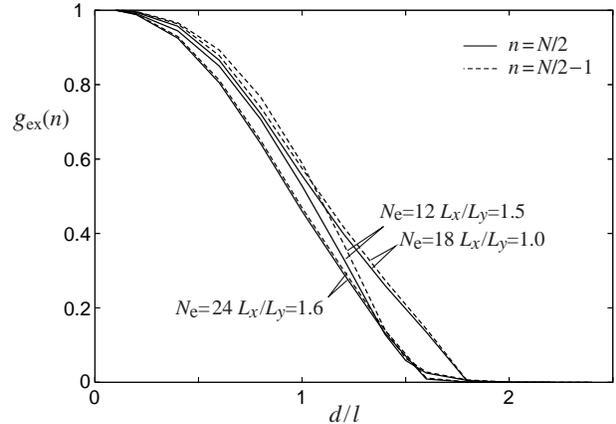}
\caption{
Exciton correlation of bilayer quantum Hall systems at
$\nu=1$. The solid line represents $g_{\rm ex}(N/2)$.
The dashed line represents $g_{\rm ex}(N/2-1)$.
}
\end{figure}

We next calculate pair correlation functions of the electrons to
see the detailed evolution of the ground-state wave function. 
The interlayer pair correlation function of the electrons is defined by 
\begin{eqnarray}
g_{12}({\mib r}) &\equiv& \frac{L_x L_y}{N_1N_2}\langle 
\Psi | \sum_{n\ m} \delta({\mib r}+{\mib R}_{1,n}
-{\mib R}_{2,m})|\Psi 
\rangle.
\end{eqnarray}
We present $\Delta g_{12}(r)$ in Fig.~3, which is defined by 
\begin{eqnarray}
\Delta g_{12}(r) &=& \frac{N_2}{L_x L_y}\int 
(g_{12}({\mib r'})-1)\delta(|{\mib r'}|-r) \ {\rm d}{\mib r'}, 
\end{eqnarray}
where ${\mib r'}$ is the two-dimensional position vector in each layer.
$\Delta g_{12}(r)$ represents the difference from the uniform correlation 
of independent electrons multiplied by $2\pi r$. 
We also show $g_{12}(r)-1$ in the inset.

At $d=0$ we find clear dips in $\Delta g_{12}(r)$ and 
$g_{12}(r)$ at $r/l  \simeq 1$ and $0$, respectively, which 
are a characteristic feature of the excitonic state made by
the binding of electrons and holes between the two layers.
The binding of one hole means the exclusion of one electron
caused by the strong interlayer Coulomb repulsion.
The increase in layer separation weakens the
Coulomb repulsion between the two layers and 
reduces $|\Delta g_{12}(r)|$ at $r/l \simeq 1$.

\begin{figure}[t]
\epsfxsize=80mm \epsffile{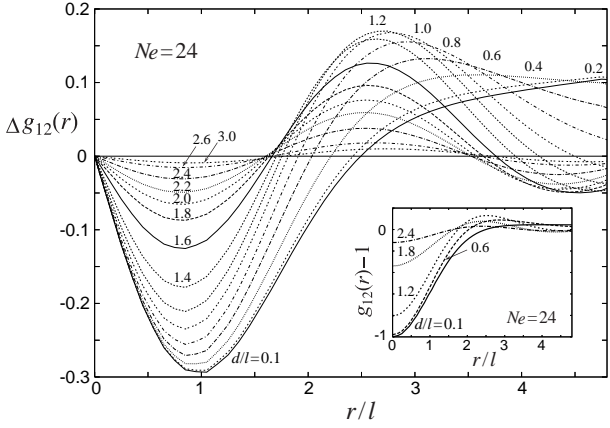}
\caption{
Interlayer pair correlation function of electrons
in ground state of bilayer quantum Hall systems at $\nu=1$. 
$N_e=24$ and $L_x/L_y=1.6$.
Inset shows $2\Delta g_{12}(r)/r$, which is essentially the same as 
$g_{12}(r)-1$.}
\end{figure}

The decrease in the interlayer correlation $|\Delta g_{12}(r)|$ 
opens space to enlarge the correlation hole in the same layer
and reduce the Coulomb energy between the electrons within the layer.
This is shown in Fig.~4, which shows the pair correlation 
functions of the electrons in the same layer defined by
\begin{eqnarray}
g_{11}({\mib r}) &\equiv& \frac{L_x L_y}{N_1(N_1-1)}\langle 
\Psi | \sum_{n m} \delta({\mib r}+{\mib R}_{1,n}-{\mib R}_{1,m})|\Psi
\rangle, 
\end{eqnarray}
\begin{eqnarray}
\Delta g_{11}(r) &=& \frac{N_1}{L_x L_y}\int  
(g_{11}({\mib r'})-1)\delta(|{\mib r'}|-r) \ {\rm d}{\mib r'} .
\end{eqnarray}
The obtained results indeed show
that the correlation hole in the same layer
at $r/l \simeq 1$ is enhanced with the increase in 
$d$ contrary to the decrease in the size of the
interlayer correlation hole in Fig.~3. 
The correlation hole in the same layer monotonically increases in size 
up to $d/l=1.8$, and then it becomes almost constant. 
The correlation function $g_{11}(r)$ for $d/l>1.8$
is almost the same as that of $\nu=1/2$ monolayer quantum Hall 
systems realized in the limit of $d=\infty$. 
This is consistent with the almost vanishing 
excitation gap and exciton correlation at $d/l>1.8$ 
shown in Figs.~1 and 2. 

\begin{figure}[t]
\epsfxsize=80mm \epsffile{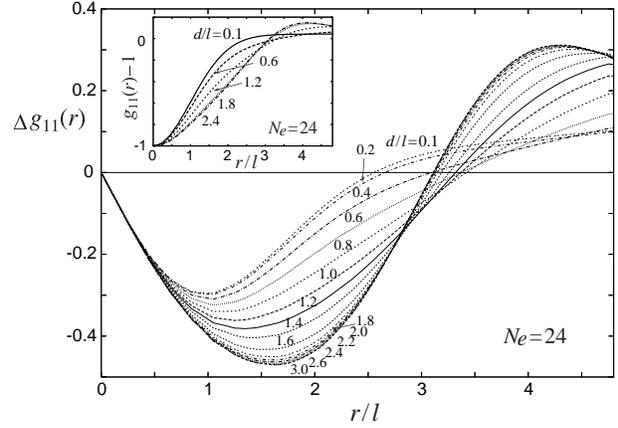}
\caption{
Intralayer pair correlation function of electrons
in ground state of bilayer quantum Hall systems at $\nu=1$. 
$N_e=24$ and $L_x/L_y=1.6$. 
The inset shows $2\Delta g_{11}(r)/r$, which is essentially the same as
$g_{11}(r)-1$.}
\end{figure}

In the $\nu=1/2$ monolayer system, the ground state is described as
a dipolar composite fermion liquid state,  
where an electron forms a bound state with a displaced hole.\cite{Read} 
The peak of $\Delta g_{11}(r)$ at $r/l\simeq 4$ shows the position
of the nearest neighbor electrons in accordance with the  
approximate mean distance between the electrons $3.54l$ 
estimated from $(L_xL_y/N_1)^{1/2}= (2\pi N/N_1)^{1/2} l$.
On the other hand, the correlation hole at $r/l\simeq 1.5$ shows
the existence of the bound hole forming the dipole. 

As the distance between the two layers decreases, the interlayer
correlation between the dipoles begins to develop. 
This is seen in Fig.~3 as the development of the correlation hole at
$r/l \simeq 1$ and a peak at $r/l \simeq 2.5$ in $\Delta
g_{12}(r)$. 
That $\Delta g_{11}(r)$ remains almost the same at $d/l > 1.6$
indicates that the identity of the dipolar composite fermion remains in
this region. 
The almost same amplitude of $\Delta g_{12}(r)$ at  
$r/l \simeq 1$ and $2.5$ for $d/l>1.6$ is also consistent with the
antiparallel interlayer correlation of dipolar composite fermions. 

With further decrease in $d/l$ from 1.6, interlayer correlation
between electrons and holes becomes stronger and begins to compete with
intralayer correlation. 
The dipolar composite fermion begins to lose its identity, and the
ground state gradually changes into a collection of interlayer
electron-hole pairs, i.e., interlayer excitons. 
The dissociation of the dipole is seen in the rapid decrease in the
size of the intralayer correlation hole at $d/l < 1.2$ in Fig.~4, and in the
broadening of the peak at $r/l \simeq 2.5$ in Fig.~3. 
On the other hand, the development of the correlation hole at $r/l
\simeq 1$ in Fig.~3 and the development of the exciton correlation in
Fig.~2 shows the formation of interlayer excitons. 
The dissociation of the dipole is also considered as continuous
weakening of the antiparallel correlation between interlayer excitons as
$d$ decreases. 
This correlation disappears at $d=0$ when the $\Psi_{1,1,1}$ state
is realized, and $\Delta g_{11}(r)$ and $\Delta g_{12}(r)$ coincide. 
In this course, half of the correlation hole in the same layer, which is a 
characteristic feature of composite fermions, 
is transferred into the other layer.

\begin{figure}[t]
\epsfxsize=80mm \epsffile{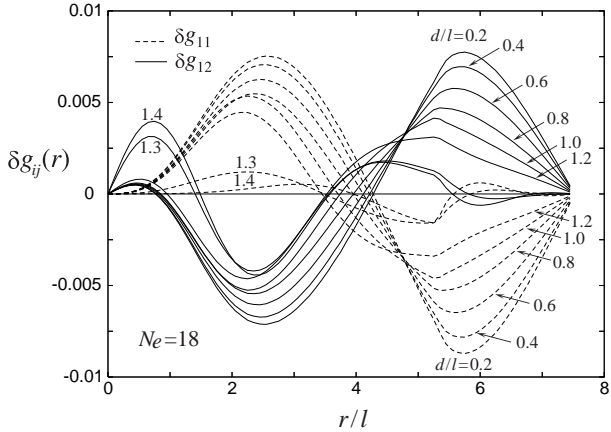}
\caption{
Change in correlation function through excitation
from ground state to first excited state.
$\nu=1$ and $N_e=18$ with $L_x/L_y=1.0$.
}
\end{figure}

\vspace{-0.4cm}
The breakdown of the composite fermions at $d/l \simeq 1.2$
affects the character of the lowest excitations, 
which is clearly shown in the 
level crossing of the excited state at $d/l\simeq 1.2$.
The change in the character of the excitation
is confirmed by the correlation functions in the
excited state. Figure 5 shows the difference in the pair correlation 
functions $g_{ij}(r)$ between the ground state
and first excited state defined by
\begin{eqnarray}
\delta g_{ij}(r) &=& \frac{N_j}{L_x L_y}\int  
(g_{ij}^{E}({\mib r'})-g_{ij}^{G}({\mib r'}))
\delta(|{\mib r'}|-r) \ {\rm d}{\mib r'} ,
\end{eqnarray}
where $g_{ij}^{G}({\mib r})$ and $g_{ij}^{E}({\mib r})$ are
the pair correlation functions 
in the ground state and the first excited state, respectively.
$\delta g_{ij}(r)$ in Fig.~5 shows that there is a discontinuous 
transition between $d/l=1.2$ and $1.3$, which supports
the level crossing in the first excited state. 

Below $d/l\simeq 1.2$, $\delta g_{ij}(r)$ have a large amplitude
at $r/l\simeq 2$ and $6$.
In this region of $d/l$, interlayer-excitons are formed and they are
weakly interacting. 
Figure 5 shows that the antiparallel correlation between excitons
realized in the ground state is weakened in the excited state. 
A small singularity at $r/l \simeq 5.3$ is due to the finite
size effects of the square unit cell.
Above $d/l\simeq 1.2$, only $\delta g_{12}(r)$ have 
large amplitude at $r/l\simeq 1$ and 2.
In this region, dipolar composite fermions are formed and there is
interlayer correlation among them. 
Figure 5 shows that the low-energy excitations are made by modifying
this interlayer correlation between dipoles. 

In summary, we have investigated the ground state 
and low-energy excitations of $\nu=1$ bilayer 
quantum Hall systems by the DMRG method. 
The obtained exciton correlations 
show that the ground state for small layer separation, $d/l < 1.6$, 
is characterized by the macroscopic coherence of excitons, $g_{\rm ex}\ne 0$, 
which is consistent with the vanishing of the Hall counterflow 
resistance\cite{CFH1,CFH2} 
and a marked enhancement in zero-bias tunneling conductance\cite{ZTNC}
at low temperatures. The excitation gap $\Delta$ rapidly decreases
with increasing $d/l$ from 1.2 to 1.6, which is also 
consistent with the clear decrease in zero bias tunneling 
conductance\cite{ZTNC} and transport 
experiments\cite{CFH1,CFH2,Wies} at finite temperatures 
around this region.

We have also studied the evolution of the ground-state wave function.
We have found a clear dip in $\Delta g_{12}(r)$ at $r/l \simeq 1$ and 
a peak at $r/l \simeq 2.5$ for $d/l > 1.2$, which is 
consistent with the interacting dipolar composite fermion picture\cite{Kim}. 
At $d/l \simeq 1.2$, the peak in $\Delta g_{12}(r)$ at 
$r/l \simeq 2.5$ becomes largest, and
the correlation functions becomes almost the same as those of 
the attractive $V_1^{12}$ model\cite{NY}.
The effective interlayer attractive interaction between the electrons 
is induced by the antiparallel correlation between dipolar composite
fermions, which can also be interpreted as the antiparallel correlation
between interlayer excitons. 
A further decrease in $d$ continuously transfers the
correlation hole to the other layer 
and the interacting interlayerexcitons gradually become a better picture
of the ground state. 
We do not find any sign of the first-order transition
through the above evolution of the wave function.
Thus, the transition from the compressible state at $d=\infty$
to the excitonic state at $d=0$ is continuous.

\acknowledgements

The present work is supported by
Grant-in-Aid No.~15740177 and No.~14540294 from JSPS.
D.Y. thanks the Aspen Center for Physics 
for their hospitality.

\end{document}